\newcommand{\al}{\mbox{$^{26}$\hspace{-0.2em}Al}}

\newcommand{\gray}{\mbox{$\gamma$-ray}}

\newcommand{\nuc}[2]{\mbox{$^{#1}$#2}}
\def\MeV{\mbox{Me\hspace{-0.1em}V}}
\def\keV{\mbox{ke\hspace{-0.1em}V}}
\def\la{\mathrel{\hbox{\rlap{\hbox{\lower4pt\hbox{$\sim$}}}\hbox{$<$}}}}
\def\ga{\mathrel{\hbox{\rlap{\hbox{\lower4pt\hbox{$\sim$}}}\hbox{$>$}}}}
\def\deg{\mbox{$^\circ$}}

\documentstyle[editedbook,epsfig,psfig,epsf]{mq}

\def\etal{{\em et al.} }

\def\cm2{cm$^2$ }
\def\se1{s$^{-1}$ }

\begin{opening}
\title{Science prospects with INTEGRAL}
\author{J\"urgen Kn\"odlseder}
\institute{Centre d'Etude Spatiale des Rayonnements, 9, avenue du 
Colonel-Roche, 31028 Toulouse Cedex 4, France.}
\end{opening}

\runningtitle{Science prospects with INTEGRAL}
\runningauthor{J\"urgen Kn\"odlseder}

\begin{document}
\vspace{-0.5cm}

\begin{abstract}
{\small
With the launch of ESA's {\em INTEGRAL} satellite in october 2002, a 
gamma-ray observatory will be placed in orbit providing a multiwavelength 
coverage from a few keV up to 10 MeV for the study of high energy phenomena 
in the universe. 
Among the major scientific objectives of {\em INTEGRAL} are the study of compact 
objects, and in particular of microquasars, stellar nucleosynthesis, high 
energy transients, particle acceleration and interaction processes, and 
galactic diffuse emission.
In this review I will expose the science prospects of {\em INTEGRAL} in the
above mentioned fields.
}
\end{abstract}

\section{The INTEGRAL observatory}

Due to continuing progress in instrumentation, the field of gamma-ray 
astronomy has become a new complementary window to the universe.
With the upcoming {\em INTEGRAL} satellite, foreseen for launch in 
october 2002, ESA provides a gamma-ray observatory to the scientific 
community that combines imaging and spectroscopic capacities in the 20 
\keV\ to 10 \MeV\ energy range.
{\em INTEGRAL} is equipped with two gamma-ray telescopes, optimised for 
high-resolution imaging (IBIS) and high-resolution spectroscopy 
(SPI), supplemented by two X-ray monitors (JEM-X) and an optical 
monitor (OMC).
With respect to precedent instruments, the {\em INTEGRAL} telescopes 
provide enhanced sensitivity together with improved angular and 
spectral resolution.

\subsection{The instruments}

IBIS is optimised for high angular resolution imaging of X- and \gray 
s in the 20 \keV\ to 10 \MeV\ domain, and is therefore called the 
imager of the {\em INTEGRAL} observatory.
Its detector system consists of two layers where the upper 
one (ISGRI) acts as hard X-ray detector that is sensitive to radiation 
in the $20-300$ \keV\ domain, while the lower one (PICsIT) absorbs the 
more energetic \gray\ photons in the 100 \keV\ - 10 \MeV\ energy range.
A tungsten mask placed at about 3.1 meters above the detector 
provides an angular resolution of about $12'$ within a fully coded 
field of view (FOV) of $9\deg$ (the partially coded FOV extends to 
$30\deg$).
The usage of CdTe and CsI(Tl) detectors allows only for a moderate 
energy resolution of $\sim6$\%, making IBIS primarily an instrument 
for studying point-like continuum sources.

SPI is optimised for high spectral resolution imaging of X- and \gray 
s in the 20 \keV\ to 8 \MeV\ energy range, and is therefore called the 
spectrometer on {\em INTEGRAL}.
Its detector system consists of 19 hexagonally shaped germanium 
detectors cooled to an operational temperature of 85 K, which provide an 
excellent spectral resolution of $\sim0.3\%$, well suited for detailed 
studies of gamma-ray lines.
A coded mask placed 1.7 meters above the detector plane provides a 
moderate angular resolution of $2.5\deg$ within a fully coded FOV of 
about $15\deg$ (the partially coded FOV extends to $\sim33\deg$).
Although SPI provides only a poor angular resolution, the usage of 
rather large mask elements makes it well suited for studying diffuse 
emission.
Since the continuum sensitivity of SPI is comparable to that of IBIS, 
it provides a perfectly complementary instrument for disentangling 
point-source and diffuse emission contributions along the galactic 
ridge (see section \ref{sec:diffusecont}).

Two identical X-ray monitors (JEM-X) extend the energy range of SPI 
and IBIS to soft X-rays ($3-35$ \keV).
They are also based on the coded mask principle and offer an angular 
resolution of $3'$, a fully coded FOV of $4.8\deg$, and a partially 
coded FOV of $13.2\deg$.
An optical monitor (OMC) provides simultaneous coverage in the 
visible $V$ band of the central part ($5\deg$) of the FOV.
Timing is possible with $1-10$ s resolution which allows linking the 
highly variable X- and \gray\ emission of compact objects to the 
visible wavelength range.

\begin{figure}[t!]
\centering
\epsfig{file=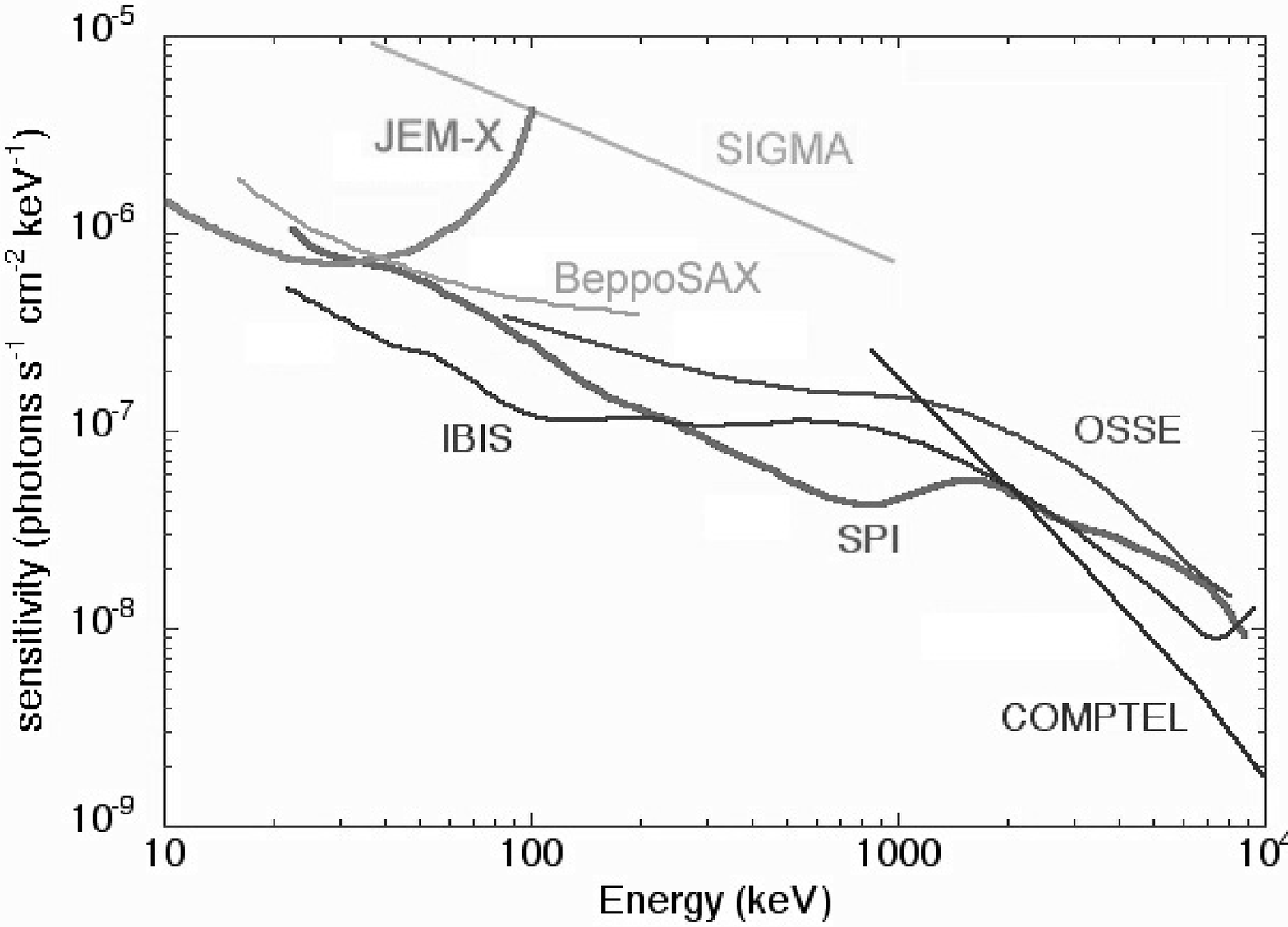,width=6.5cm}
\hfill
\epsfig{file=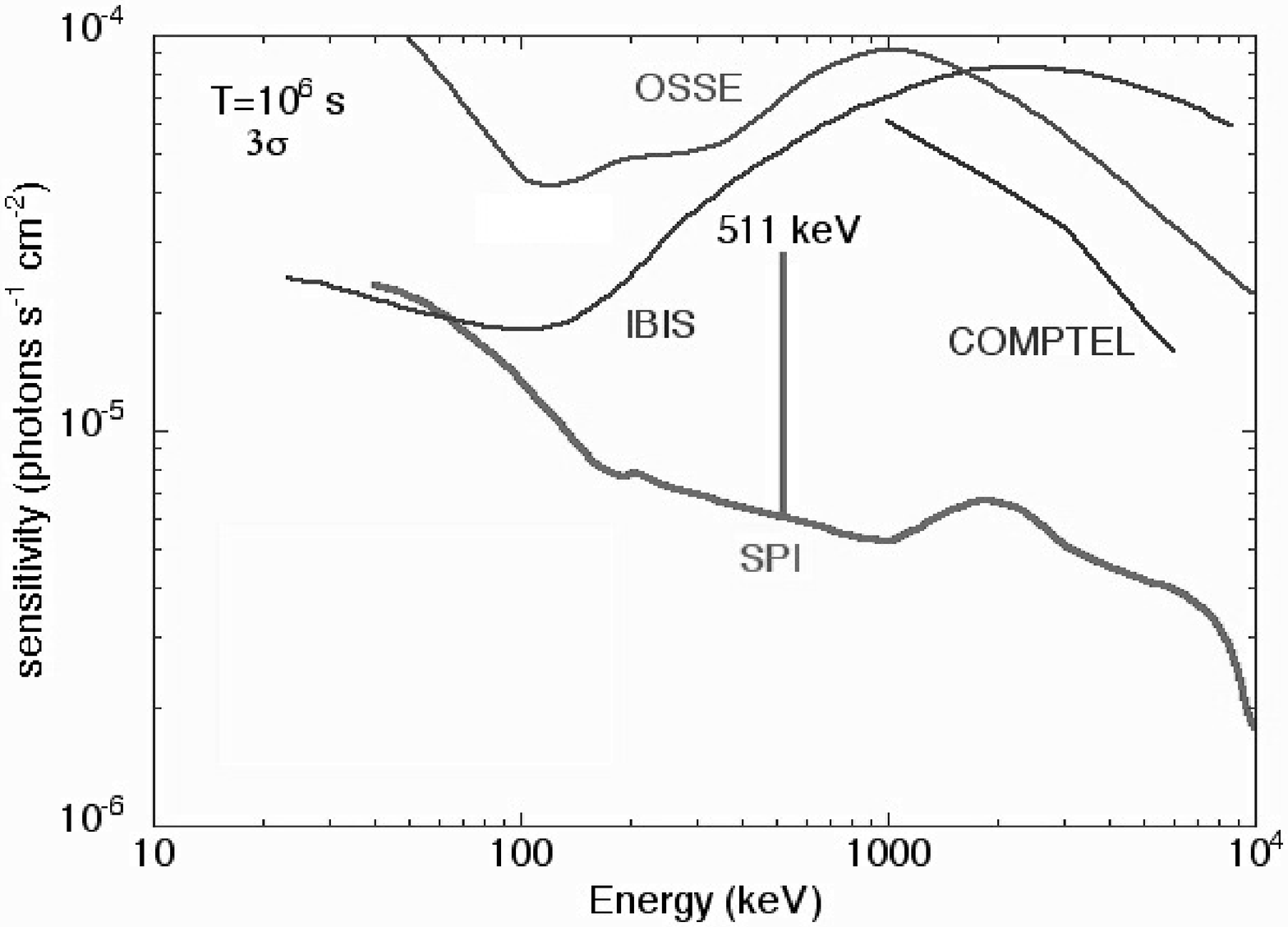,width=6.5cm}
\caption{Comparison of {\em INTEGRAL} continuum (left) and line 
(right) sensitivities to the performances of precedent instruments.}
\label{fig:sensitivity}
\end{figure}

Figure \ref{fig:sensitivity} compares the expected sensitivities of 
IBIS, SPI, and JEM-X to those of precedent hard X-ray and gamma-ray 
telescopes.
Obviously, {\em INTEGRAL} will provide unprecedented continuum 
sensitivity in the energy range from a few tens of \keV\ up to a few 
\MeV.
In particular, IBIS has a much better angular resolution than the OSSE 
telescope aboard {\em CGRO} ($12'$ versus $4\deg$), allowing precise 
localisation (and hence identification) of gamma-ray sources (SIGMA 
had a similar angular resolution, yet its limited sensitivity did allow 
only the detection of bright sources).
Concerning gamma-ray lines, SPI is expected to provide a major 
breakthrough with respect to precedent missions.
Not only is the sensitivity improved by an important factor, but the 
combination with the extremely high energy resolution will allow for 
the first time detailed line profile studies in the gamma-ray domain.

\subsection{Mission profile}

{\em INTEGRAL} is an observatory-type mission with a nominal lifetime 
of 2 years and a (plausible) extension up to 5 years.
Most of the observing time ($65-75\%$) is awarded to the scientific 
community through a standard AO process, the remaining time is 
reserved as guaranteed time for the PI institutes, Russia and NASA 
for their contributions to the program, and to ESA/ISOC and the 
mission scientists.
During the guaranteed time, the so-called {\em Core Program} is 
executed which is composed of
\begin{itemize}
\item weakly galactic plane scans (GPS) with the primary aim of detecting 
      transient sources,
\item the galactic centre deep exposure (GCDE) which aims in studying 
      the source population and diffuse emission distributions in the 
      central galactic radian,
\item target of opportunity (ToO) follow-up observations,
\item and a deep exposure of the Vela region during the first year of 
      the mission.
\end{itemize}

\section{Legacy}
 
Although various \gray\ telescopes have preceded the {\em INTEGRAL} 
observatory in its quest for unveiling the high-energy universe,
two missions have particularly influenced the design and the performance 
parameters of the {\em INTEGRAL} telescopes:
the French/Russian telescope SIGMA on the {\em GRANAT} satellite and the 
telescopes COMPTEL and OSSE on NASA's {\em Compton Gamma-Ray Observatory} 
({\em CGRO}).
 
{\em GRANAT} has been launched in december 1989 from Baikonour, 
Kazakhstan, using a Proton rocket, and has successfully been operated 
during 8 years ({\em INTEGRAL} will be lift into orbit in an 
identical scenario).
The SIGMA telescope was based on the coded mask principle and covered 
the hard X-ray to soft \gray\ band from 35 \keV\ - 1.3 \MeV.
With an angular resolution of $13'$ and a $4.7\deg \times 4.3\deg$ 
FOV, SIGMA detected more than 30 sources which are all associated to 
compact objects, such as 
persistent black hole candidates (BHCs),
X-ray novae,
Type I X-ray busters,
accreting and isolated pulsars, and
Active Galactic Nuclei (AGN).
Most of these sources are highly variable, and a substantial fraction 
only shows transient emission.
Consequently, regular repeated observations are needed to study their 
time variability, and search scans are required to unveil new 
transient sources, which largely motivated the inclusion of the GPS in 
the {\em INTEGRAL} mission profile.
In addition, SIGMA observations showed that the inner Galaxy is densely 
populated with sources, and a good angular resolution (as implemented 
in IBIS) is mandatory to avoid source confusion, and to allow for 
source counterpart identification.
 
{\em CRGO} has been launched in april 1991 by NASA's space shuttle, and 
remained in orbit during 9 years after which the satellite was 
deorbited in a spectacular manoeuvre.
COMPTEL consisted of a Compton telescope and covered the energy range 
from 750 \keV\ - 30 \MeV\ with a moderate angular ($4\deg$) and 
energy ($9\%$) resolution.
Amongst its most important achievements were the first survey of the 
entire sky in this energy domain, providing maps of diffuse \gray\ 
line (1.809 \MeV) and continuum emission.
In addition, 
pulsars, 
AGNs, 
BHCs, 
Gamma-Ray Bursts (GRBs), and 
supernova remnants (SNRs) have been detected \cite{schoenfelder00}.
The OSSE telescope consisted of collimated ($3.8\deg \times 
11.4\deg$) scintillator crystals with moderate ($7\%$) energy 
resolution, and extended the energy band covered by {\em CRGO} to lower 
energies (50 \keV\ - 10 \MeV).
OSSE provided the first map of the 511 \keV\ positron annihilation 
radiation from the central galactic radian, but it also detected 
\gray\ lines in the nearby SN~1987A supernova remnant.
It also observed numerous galactic and extragalactic compact objects 
and provided a precise measurement of the diffuse hard X-/soft \gray\ 
continuum emission of the galactic ridge.
Both COMPTEL and OSSE have revealed that diffuse \gray\ emission 
extends over angular scales of several degrees, which set the 
resolution scale and FOV of SPI.
These telescopes also showed that diffuse \gray\ emission extends over 
the entire galactic plane with pronounced enhancements towards the central 
radian, which motivated the inclusion of the GPS and GCDE in the 
{\em INTEGRAL} core program in order to build-up a substantial exposure 
of these important regions over the mission lifetime.

\section{Compact objects}

\subsection{Black-hole candidates, X-ray novae, microquasars}

The class of black hole candidates (BHC) divides into persistent sources 
and soft X-ray transients, also called X-ray novae.
The persistent sources are all high-mass X-ray binaries (HMXB), 
the prototype being Cyg X-1, while X-ray novae are low-mass X-ray 
binaries (LMXB).
X-ray novae typically reach their maximum luminosity within a few days, 
from where on the luminosity regularly declines on a few weeks timescale.
Recurrence intervals for X-ray novae are in the range $1-50$ yr, 
which suggests an underlying population between 200-1000 black-hole 
binaries within the Galaxy \cite{tanaka96,chen97}.
Due to its improved sensitivity with respect to precedent 
instruments, IBIS has the potential to detect them all, improving 
considerably our knowledge about their distribution and 
characteristics.

Probably the most characteristic spectral signature of BHCs is the 
hard power-law tail that extends to energies $\ga 200$ \keV.
The tail is generally interpreted as evidence for Comptonisation of 
soft photons associated with an accretion disk by a semi-relativistic 
plasma.
The tail often obeys an exponential cut-off that shifts to lower 
energies with increasing luminosity, probably as result of enhanced 
Compton cooling.
Yet, actual models have difficulties to explain the extended tails 
observed in Cyg X-1 or GRO J0422+32 that reach up to MeV energies 
\cite{mcconnell94,vandijk95}.
Here, better observational data is needed in the few 100 \keV\ to \MeV\ 
region that allows a better determination of the spectral shape.

There is increasing evidence that the hard X-ray emission produced by 
accretion onto stellar mass black holes is always associated to the 
production of relativistic jets \cite{mirabel99}.
Early radio observations with the VLA interferometer of NRAO of 
several SIGMA sources were actually very successful and lead to the 
identification of the radio counterpart of X-ray Nova Oph 1993,
the finding of radio jets in two persistent hard X-ray sources
(1E~1740.7-2942 and GRS~1758-258), and the discovery of GRS~1915+105 
as the first superluminal radio source in the Galaxy.
This illustrates the importance of scheduling contemporaneous 
X-/\gray\ and radio observations, a possibility that is offered by 
{\em INTEGRAL} through the GPS and GCDE observations that are well 
planned in advance and provide the highest likelihood in detecting 
new transient sources.

The study of outburst phenomena is important since it generally reflects 
a change in the accretion rate, hence it provides an excellent laboratory 
to study accretion physics and the formation mechanism of relativistic 
jets.
Multi-wavelength observations of the microquasar GRS~1915+105, for 
example, have been interpreted as ejection of the plasma corona in form 
of collimated jets as result of a shock that has been formed due to the 
refill of the accretion disk \cite{mirabel01}.
Yet, the matter content in the jets (normal plasma made of electrons 
and protons or a pair plasma made of electrons and positrons, or a mixture 
of both) is still an open question, which may be resolved by the 
detection of annihilation features during jet ejection.
Transient line features that might be indicative of pair annihilation have 
been reported in 1E~1740.7-2942, Nova Muscae, and GRO~1655-40
\cite{bouchet91,goldwurm92}, yet contemporaneous observations by OSSE 
and SIGMA of an outburst of 1E~1740.7-2942 gave contradictory results 
\cite{jung95}.
Hopefully, with its good sensitivity and spectral resolution, {\em 
INTEGRAL} will clarify the situation.

\subsection{Type I X-ray bursters}

Type I X-ray bursters are interpreted as thermonuclear explosions at the 
surface of weakly magnetised neutron stars, and are associated with 
LMXBs.
They typically occur at intervals of hours to several days, and show 
characteristic spectral changes which are interpreted as photospheric 
temperature changes during the outburst \cite{hoffman78}.
In particular, the observations of Eddington-limited bursts provide 
important insight into parameters of neutron star binaries, such as 
estimates of the distance, neutron star radii, average luminosities, and 
accretion rates \cite{cocchi01}.
With the improved sensitivity of IBIS, {\em INTEGRAL} will increase 
the sample of known X-ray bursters, and will allow detailed studies of 
their characteristics and their galactic distribution.
In particular, the new class of X-ray bursters that has been revealed 
recent by {\em BeppoSAX} observations \cite{cocchi01} is in the focus 
of {\em INTEGRAL} observations, which should improve our knowledge 
about the proportion of black-hole to neutron star systems in our 
Galaxy.

\subsection{Pulsars}

Pulsars are highly magnetised rotating neutron stars that spin-down 
due to particle acceleration processes in their magnetosphere, which 
are most directly probed in the X- and \gray\ domain \cite{harding99}.
However, it is not yet clear in which region of the magnetosphere 
electrons (or positrons) are accelerated and whether processes like 
magnetic pair production and photon splitting are relevant or not.
As prototype of an isolated pulsar, the Crab pulsar shows a 
pulse shape transition in the hard X-ray to soft \gray\ domain which 
possibly indicates the presence of different emission components
\cite{massaro01}.
In general, turn-overs in the hard X-ray domain are expected, which 
potentially help to distinguish between the different pulsar models 
\cite{harding99}.
With the good sensitivity in this transition region, {\em INTEGRAL} 
will shed new light on the origin of the different pulse components.

In the X-ray domain, most pulsars are accreting binaries where the 
mass flow is confined at the magnetic poles.
Shocks in the accreting column heat the matter up to \gray\ 
temperatures, which results in highly variable sources of hard 
X-/soft \gray\ photons.
In the extreme magnetic fields ($B \ga 10^{12}$ G) that characterise 
most accreting X-ray pulsars, the energy of the electrons is 
quantified in Landau levels, leading to resonant scattering that 
induces cyclotron absorption line features at energies
$E = 11.6 \times n \times B_{12}$ \keV, where $n$ is a positive integer 
number and $B_{12}$ is the magnetic field strength in units of 
$10^{12}$ Gauss.
Consequently, the determination of the cyclotron line energies allows 
for a direct measurement of the magnetic field strength.
Here {\em INTEGRAL} is particularly powerful since it fully covers 
the energy domain of cyclotron features (many precedent instruments 
were not sensitive enough in the hard X-ray band to reliably 
determine multiple harmonics of the same cyclotron feature).
In particular, thanks to its high spectral resolution, SPI will probe
the geometrical conditions of the emission by detailed cyclotron line 
shape measurements \cite{kretschmar01}.

\subsection{Active Galactic Nuclei}

Active Galactic Nuclei (AGN) are believed to consist of an accreting
supermassive black hole, where the central region is populated by a hot 
electron corona surrounded by an extended molecular cloud torus.
Depending on the viewing angle, different types of AGNs may be 
observed \cite{urry95}.
Most AGNs detected above $\sim50$ \keV\ are either
blazars (radio-loud AGNs viewed along the relativistic jet), or
radio-quiet Seyferts (AGNs without strong jet emission).

Blazars show important flaring activity on timescales of days and 
less, and obey a characteristic spectral turnover at \MeV\ energies.
In the X-/\gray\ domain, blazar emission is generally interpreted as 
inverse Compton radiation generated by relativistic electrons on soft 
photons.
At soft X-ray energies, a synchrotron component from relativistic electrons 
and eventually a thermal accretion disk component may contribute.
The most pertinent questions on which {\em INTEGRAL} may shed new light is 
the nature of the primary accelerated particles in the jet (leptons 
or hadrons) and the origin of the soft photons in leptonic jets 
(synchrotron self compton or external Compton scattering).
Eventually, the detection of positron annihilation features by 
{\em INTEGRAL} may reveal information about the jet composition.

Radio-quiet Seyferts are generally only detected up to a few 100 \keV, 
showing a thermal Comptonisation spectrum with an exponential cut-off.
The emission source is typically located above the surface of the 
accretion disk, by which it is Compton reflected.
Yet, the origin of the incident primary radiation is much less 
understood.
Key observables, that will be provided by {\em INTEGRAL}, are the spectral 
shape at hard X-ray/soft \gray\ energies (break or cut-off) to 
determine the origin of the radiation.
In addition, the  contribution of non-thermal processes, such as jet 
emission, can be constrained.
High-energy observations can determine the fraction of non-thermal 
processes, such as the acceleration of electrons (or 
electron-positron pairs) to relativistic energies.
This results in a high-energy tail that is either due to Compton 
scattering by relativistic electrons or due to electron-positron 
annihilation from pair cascades.

{\em INTEGRAL} may also play a fundamental role in establishing the 
evolutionary link between galaxy mergers and AGNs.
It is believed that galaxy mergers and interactions lead to 
infrared luminous galaxies which harbor AGN that are fueled and 
uncovered as the dust settles.
In the optical domain, dust obscuration hides the potential AGN which 
prevents the verification of this scenario.
In contrast, due to the penetrating nature of \gray s, {\em INTEGRAL} 
observations may reveal the hidden AGN in infrared luminous 
galaxies.

\subsection{Gamma-ray bursts}

Owing to the large FOVs of IBIS and SPI, about 12 \gray\ bursts per 
year are expected to be detected by the instruments on {\em INTEGRAL}.
Although no onboard burst localisation is foreseen, the realtime 
telemetry allows for a rapid burst localisation with arcmin 
precision within a typical alert time of $\sim30$ seconds after burst 
occurrence.
For this purpose, a special Integral Burst Alert System (IBAS) has been 
installed at the Integral Science Data Centre (ISDC) \cite{mereghetti99}.
If the burst occurs in the (smaller) FOV of the optical monitor 
aboard {\em INTEGRAL}, a rapid reconfiguration procedure of the OMC 
has been foreseen in order to obtain images in the visible band 
around the estimated burst location.
Thus {\em INTEGRAL} may provide itself optical counterpart identifications,
although the small OMC FOV ($5\deg \times 5\deg$) make such events probably 
rather rare.

With an effective area of $\sim3000$ cm$^2$, the SPI anticoincidence 
shield, operating above $\la100$ \keV, provides an excellent large area 
detector that may also be used for GRB detection.
It is expected that it may detect several hundred bursts per year, 
which will be dated to 50 ms accuracy \cite{lichti00}.
By adding this timing information to that of other satellites in the 
interplanetary network of GRB detectors, it is expected to localise 
GRBs to within an area of 8 square arcminutes \cite{hurley01}.

\section{Diffuse emission}

\subsection{Galactic continuum emission}
\label{sec:diffusecont}

Measurements of the galactic continuum emission in the hard 
X-ray/soft \gray\ band is inherently difficult because of the presence 
of numerous transients and variable discrete sources in the galactic plane, 
and the fact that X-/\gray\ telescopes either have large fields 
of view and no imaging capabilities, or have imaging capability but no 
diffuse emission sensitivity.
{\em INTEGRAL}, with its complement instruments IBIS (high angular 
resolution) and SPI (good diffuse emission sensitivity), provides an 
unprecedented combination that finally will allow for an accurate 
separation of point source and diffuse emission components.

In the continuum energy range where {\em INTEGRAL} is most sensitive (few 
tens of \keV\ to few \MeV), the galactic ridge emission is composed of
(1) transient discrete sources,
(2) the positron annihilation line and 3 photon continuum, 
(3) a soft \gray\ component ($\la 300$ \keV) of uncertain origin, and
(4) a hard \gray\ component ($\ga 500$ \keV) which is likely due to 
interaction of cosmic-rays with the interstellar medium  \cite{valinia01}.
The origin of the soft \gray\ component is certainly one of the key 
questions that will be addressed by {\em INTEGRAL}.
If it is made of hard, discrete sources, IBIS should be able to 
unveil at least some of them.
However, it may turn out that the soft \gray\ component is 
intrinsically diffuse, likely due to low-energy cosmic ray electrons 
that undergo Bremsstrahlung while encountering the interstellar medium.

\subsection{Positron annihilation signatures}

The 511 \keV\ gamma-ray line due to annihilation of positrons and 
electrons in the interstellar medium has been observed by numerous 
instruments \cite{harris98}.
At least two galactic emission components have been identified so far:
an extended bulge component and a disk component.
Indications of a third component situated above the galactic centre 
have been reported, yet still needs confirmation.

The galactic disk component may be explained by radioactive positron 
emitters, such as \nuc{26}{Al}, \nuc{44}{Sc}, \nuc{56}{Co}, and 
\nuc{22}{Na}.
The origin of the galactic bulge component is much less clear, and 
will be one of the key-questions addressed by {\em INTEGRAL}.
SPI will provide a detailed map of 511 \keV\ emission from the Galaxy.
Using this map, the morphology of the galactic bulge can be studied in 
detail, and the question on the contribution of point sources to the 
galactic bulge emission can be addressed.
In particular, the ratio between bulge and disk emission, which is only 
poorly constrained by existing data, will be measured more precisely, 
allowing for more stringent conclusions about the positron sources of both 
components.
The 511 \keV\ map will also answer the question about the reality of the
positive latitude enhancement, which may provide interesting 
clues on the activity near to the galactic nucleus.

The 511 \keV\ line shape carries valuable information about the annihilation 
environment which will be explored by SPI.
The dominant annihilation mechanism sensitively depends on the temperature, 
the density, and the ionisation fraction of the medium, and the measurement 
of the 511 \keV\ line width allows the determination of the 
annihilation conditions \cite{guessoum91}.
Observations of a moderately broadened 511 \keV\ line towards the 
galactic centre indicate that annihilation in the bulge mainly occurs
in the warm neutral or ionised interstellar medium \cite{harris98}.
By making spatially resolved line shape measurements, SPI will allow 
to extend such studies to the entire galactic plane, complementing 
our view of galactic annihilation processes.

With its good continuum sensitivity, SPI will also be able to detect 
the galactic positronium continuum emission below 511 \keV.
The intensity of this component with respect to that of the 511 \keV\ 
line carries complementary information about the fraction $f$ of 
annihilations via positronium formation, probing the thermodynamical and 
ionisation state of the annihilation environment \cite{guessoum91}.

\subsection{Galactic 1.809 \MeV\ line emission}

The COMPTEL telescope aboard {\em CGRO} has provided the first all-sky map 
of the distribution of the radioactive \al\ isotope that, with a lifetime of 
$\sim10^6$ years, is an unambiguous proof that nucleosynthesis is still 
active in our Galaxy \cite{oberlack96}.
The 1.809 \MeV\ \gray\ dacay line map shows the galactic disk as the most 
prominent emission feature, demonstrating that \al\ production is clearly a 
galaxywide phenomenon.
The observed intensity profile along the galactic plane 
reveals asymmetries and localised emission enhancements, 
characteristic for a massive star population that follows the 
galactic spiral structure.
By refining the knowledge about the 1.809 \MeV\ emission distribution,
SPI will provide a unique view on the star formation activity in our Galaxy.

A precise determination of the 1.809 \MeV\ latitude profile by SPI
will provide important information about the dynamics and the mixing 
of \al\ ejecta within the interstellar medium.
High velocity \al\ has been suggested by measurements of a broadened 
1.809 \MeV\ line by the {\em GRIS} spectrometer \cite{naya96}, although 
this observation is at some point at odds with the earlier 
observation of a narrow line by {\em HEAO 3} \cite{mahoney84}.
In any case, the propagation of \al\ away from its origin should lead 
to a latitude broadening with respect to the scale height of the 
source population, and the observation of this broadening may allow 
the study of galactic outflows and the mass transfer between disk and 
halo of the Galaxy.
The excellent energy resolution of SPI will easily allow to decide 
whether the 1.809 \MeV\ line is broadened or not, and the improved
angular resolution and sensitivity with respect to COMPTEL will 
allow to determine the scale height of the galactic \al\ distribution 
much more precisely.

The energy resolution of SPI of $\sim2.5$ \keV\ at 1.809 \MeV\ 
converts into a velocity resolution of $\sim400$ km s$^{-1}$, allowing 
for line centroid determinations of the order of $50$ km s$^{-1}$ for 
bright emission features.
Thus, in the case of an intrinsically narrow 1.809 \MeV\ line, line 
shifts due to galactic rotation should be measurable by 
SPI \cite{gehrels96}.
Although this objective figures certainly among the most ambitious 
goals of SPI observations, a coarse distance determination of 1.809 \MeV\ 
emission features based on the galactic rotation curve seems in 
principle possible.

\subsection{Galaxy intracluster emission}
 
Recent radio, EUV and X-ray observations suggest that clusters of 
galaxies contain large populations of non-thermal relativistic and/or 
superthermal particles \cite{goldoni01}.
These particles may be produced by acceleration in cluster merger 
shocks, AGNs, and supernova explosions in cluster galaxies.
$20-100$ \keV\ excess emission has been reported by {\em BeppoSAX} and 
{\em RXTE} at 
a level of about 2 mCrab from the Coma cluster, yet source confusion, 
in particular with the nearby Seyfert 1 galaxy X Comae, make the precise 
flux level uncertain.
With its high angular resolution and good sensitivity, IBIS is 
optimally suited to disentangle the various source contributions, and 
to allow a precise determination of the non-thermal intracluster 
emission.
Hard X-ray observations of intracluster gas directly yield the mean 
strength of the magnetic fields and energy density of relativistic 
electrons.

\section{Nucleosynthesis}

\subsection{Supernovae}

Supernovae are the most prolific nucleosynthesis sites in the universe, 
producing a large variety of chemical elements that are ejected into 
the interstellar medium by the explosion.
Among those are radioactive isotopes (such as \nuc{56}{Ni}) that decay 
under gamma-ray line emission with lifetimes that are sufficiently long 
to allow escape in regions that are transparent to gamma-rays.

Type Ia events are easier to observe than the other
supernova classes because they produce an order of magnitude more radioactive 
\nuc{56}{Ni} than the other types, and because they expand rapidly enough to 
allow the gamma-rays to escape before all the fresh radioactivity has 
decayed.
From the SPI sensitivity and the observed type Ia supernova rates 
together with standard models of type Ia nucleosynthesis, one may 
estimate the maximum detectable distance for a type Ia event to about 
$15$ Mpc and the event frequency to one event each 5 years \cite{timmes97}.
Hence, during an extended mission lifetime of 5 years, SPI has 
statistically spoken the chance to detect one such event, which would
provide important clues on the nature of the explosion mechanism.

\subsection{Supernova remnants}

Unveiling historic supernovae by searching for the \gray\ lines from 
the radioactive decay of \nuc{44}{Ti} is a further key science 
objective of {\em INTEGRAL}.
The proof of principle has been achieved by the observation of
1.157 \MeV\ \gray\ line emission from the 320 years old Cas A supernova 
remnant using the COMPTEL telescope \cite{iyudin94}.
Evidence for another galactic \nuc{44}{Ti} source (RX J0852.0-4622) has 
been found in the Vela region where no young supernova remnant was known 
before \cite{iyudin98}.

Given the marginal detection of the 1.157 \MeV\ line from 
RX J0852.0-4622, a confirmation by SPI will be crucial for the further 
understanding of this object.
\nuc{44}{Ti} line-profile measurements will provide complementary 
information on the expansion velocity and dynamics of the innermost 
layers of the supernova ejecta.
The regular galactic plane scans and the deep exposure of the central 
radian that {\em INTEGRAL} will perform during the core program will 
lead to a substantial build-up of exposure time along the 
galactic plane, enabling the detection of further hidden young galactic 
supernova remnants through \nuc{44}{Ti} decay.
The observed supernova statistics may then set interesting constraints on 
the galactic supernova rate and the \nuc{44}{Ti} progenitors.
Indeed, actual observations already indicate that some of the galactic 
\nuc{44}{Ca} may have been produced by a rare type of supernova (e.g. 
Helium white dwarf detonations) which produces very large amounts of 
\nuc{44}{Ti} \cite{the99}.

\section{Conclusions}

The list of {\em INTEGRAL} science prospects given above is by far not 
complete.
In addition, studies of
supernova remnants,
Wolf-Rayet and Be stars,
unidentified EGRET sources,
cosmic background,
\nuc{60}{Fe} and nova nucleosynthesis,
and the observation of nuclear interaction and neutron capture lines
are on the menu of {\em INTEGRAL} science prospects, illustrating the 
richness of sources and the variety of information that can be obtained 
in the hard X- to soft \gray\ domain \cite{knoedlseder01}.
The combination of high angular resolution (IBIS) with excellent 
spectral resolution (SPI), covering the X-ray (JEM-X) to \gray\ band, 
provides a unique instrument configuration that has the potential to 
explore the high-energy universe far beyond the established horizon.


\end{document}